\documentstyle[11pt,aaspp4]{article}
\begin{document}
\title{THE DISTRIBUTION OF DARK MATTER \\
IN A RINGED GALAXY }

\author{
%A. C. Quillen\altaffilmark{1}$^,$\altaffilmark{2}$^,$\altaffilmark{3}$^,$\altaffilmark{4}  
A. C. Quillen\altaffilmark{1}$^,$\altaffilmark{2}$^,$\altaffilmark{3}  \&  
J. A. Frogel${^{2,}}$\altaffilmark{4}$^,$\altaffilmark{5}
}

\altaffiltext{1}{University of Arizona, Steward Observatory, Tucson, AZ 85721}
\altaffiltext{2}{Astronomy Department, Ohio State University, 174 W. 18th Ave., 
    Columbus, OH 43210}
%\altaffiltext{3}{E-mail: quillen@payne.mps.ohio-state.edu}
\altaffiltext{3}{E-mail: aquillen@as.arizona.edu}
\altaffiltext{4}{Visiting Astronomer at Cerro Tololo Interamerican Observatories}
\altaffiltext{5}{Department of Physics, University of Durham, Durham, England}

\begin{abstract}
Outer rings are located at the greatest distance from the galaxy center 
of any feature resonant with a bar.  Because of their large scale,
their morphology is sensitive to 
the distribution of the dark matter in the galaxy.
We introduce here how study of these rings can constrain
the mass-to-light ratio of the bar, and so the percentage of dark
matter in the center of these galaxies. 

We  compare periodic orbits integrated in the ringed galaxy NGC~6782 
near the outer Lindblad resonance to the shape of the outer ring.
The non-axisymmetric component of the potential resulting
from the bar is derived from a near-infrared image of the galaxy.  
The axisymmetric component is derived assuming a flat rotation curve.  
We find that the pinched non-self-intersecting 
periodic orbits are more elongated for higher bar mass-to-light ratios
and faster bars.
The inferred mass-to-light ratio of the bar depends on the assumed
inclination of the galaxy.  With an assumed  galaxy inclination of $i=41^\circ$,
for the orbits to be consistent with the observed ring morphology
the mass-to-light ratio of the bar must be high, greater than $70\%$ of
a maximal disk value.    For $i=45^\circ$, the mass-to-light ratio
of the bar is $75\pm 15\%$ of the maximal disk value. 

Since the velocity field of these rings
can be used to constrain the galaxy inclination as well as 
which periodic orbit is represented in the
ring, further study will yield tighter constraints on the mass-to-light
ratio of the bar.   If a near maximal disk
value for the bar is required, 
then either there would be little dark matter 
within the bar, or the dark matter contained in the disk of the galaxy
would be non-axisymmetric and would rotate with the bar. 

\end{abstract}


\section {INTRODUCTION}
Outer rings are located at the greatest distance from the 
galaxy center of any feature resonant with a bar.  
Because of their great distance from the galaxy center,
their morphology should be sensitive to 
the distribution of the dark matter in the galaxy.
Since they are resonant with the bar they are also sensitive 
to the mass-to-light ratio of the bar.  
These rings provide a unique opportunity to constrain
the mass-to-light ratio of the luminous stellar matter 
in a galaxy, thus telling us about the dark matter distribution.
%By measuring the gravitational force of a non-axisymmetric component, the bar,
%which is necessarily a disk component, 
%we can constrain the disk mass-to-light ratio in a way that is independent
%of any assumption about the dark matter distribution.
Although the morphology of outer rings has been used to 
constrain bar pattern speeds
(\cite{byr94}), it has not yet been used to constrain the dark matter
distribution.

Disk mass-to-light ratios have been almost exclusively measured 
by axisymmetric fits to observed rotation curves 
(e.g. \cite{ken87a}, \cite{ken87b}, \cite{bro97}, \cite{beg91}, \cite{sac97b}) 
and are commonly done by requiring the disk to be as
massive as possible, as in the ``maximal disk'' model.
Disk mass-to-light ratios determined in this manner are model dependent
(see \cite{sac97a}) and are affected by a variety of assumptions such as
the dark matter halo profile assumed and the bulge/disk decomposition method.
In this paper,
by measuring the gravitational force of a non-axisymmetric component, the bar,
which is necessarily a disk component, 
we can constrain the disk mass-to-light ratio in a way that is independent
of the radial distribution of the dark matter,
and does not require the assumption (of the ``maximal disk'' model) that
the disk is as massive as possible within the limits set by the rotation curve.

% Classification
Three classes of rings can be seen in normal barred galaxies.
Nuclear rings located inside the bar, inner rings which envelop the bar, 
and outer rings which surround the bar.  
The simulations of \cite{sch81} first demonstrated that outer rings 
develop in the ISM near the Outer Lindblad Resonance (OLR) of the bar.
Outer rings are classified (following \cite{but91}) as R1-type or R2-type rings,
where R2-type rings are oval and aligned parallel to the bar and  
R1-type rings resemble an oval with dimples at the points where the ring
is closest to the bar.
When the rings contain spiral structure, they are classified as pseudo-rings
and are denoted R1$'$ or R2$'$ rings. 
The two types of rings have morphology closed related to 
the two families of closed orbits near the OLR (e. g. see  \cite{con89} or
\cite{kal91}).
The bar in these galaxies causes many stellar orbits in the 
plane of the galaxy to intersect themselves.
Because gas can shock, it cannot remain in these orbits, so it 
collects in orbits that are near the periodic orbit families which are not
self-intersecting (\cite{sch81}). 
For an excellent review on the properties of ringed galaxies see \cite{but96}.

In this paper we examine the sensitivity of the ring shape of an R1$'$ ringed
galaxy to the mass-to-light ratio of the bar.  
The non-axisymmetric component of the gravitational potential is derived
from a $J$ band image of the galaxy.  The morphology of the ring is
compared to periodic orbits near the OLR for different bar mass-to-light ratios.
Using this comparison we place limits on the mass-to-light ratio of the bar, 
and so on the distribution of the dark matter in the galaxy.

\section{THE IMAGES OF NGC~6782}
% The data
NGC~6782 is the best example of an R1-type ringed galaxy found 
to date in the OSU galaxy
survey.  In a catalog of southern ringed galaxies (\cite{but95}) the galaxy
is classified as R1$'$SB(r)0/a.  
\cite{but95} shows this galaxy as a nice example of an R1$'$ ringed galaxy
and  demonstrates that 
the outer ring is prominent in a $B-I$ color map and so is quite blue.
\cite{byr94} show NGC~6782 as a example of a galaxy
that resembles their slow pattern speed simulations.  
The R1$'$ ring in NGC~6782 has a radius of $\sim 60''$ which
corresponds to 15kpc 
assuming a distance to the galaxy of 50 Mpc ($H_0 = 75$ km s$^{-1}$ Mpc$^{-1}$).

The galaxy was observed in the
near infrared $J,H$ and $K$ bands and in the visible $B,V$ and $R$ bands.
These data are a preliminary part of a survey being carried
out at the Ohio State University of $\sim 220$
galaxies (\cite{fro96}).
The survey's goal is to produce a library of photometrically
calibrated images of late-type galaxies from $0.4$ to $2.2 \mu$m.
For notes on the observation and reduction techniques see
\cite{pog97}, or for individual examples \cite{qui94}, and \cite{qui95}.
All the images were observed 
at the Cerro Tololo Interamerican Observatories.
The $BVR$ images were observed 
at the 0.9m telescope on 1995 October 25 using the Tek\#2
$1024\times1024$ pixel CCD with a spatial scale of $0.\hskip-2pt''40$/pixel.
Total on source exposure times were 30, 15 and 10 minutes for $B,V$ and
$R$ respectively.  
The $JHK$ images were observed at the 1.5m telescope 
on 1995 October 31 and 1995 Nov 2 using the NICMOS 3 $256\times256$
pixel infrared array with a spatial scale of $1.\hskip-2pt''16$/pixel.
Total on source exposure times were 15 minutes at $J$ and $H$ and
$29$ minutes at $K$ band.
The infrared images were observed during photometric conditions and were
calibrated on the CTIO/CIT system using standard stars listed by 
Carter \& Meadows 1995.
% Brian Carter at SAAO and Vicki Meadows at AAO 
The optical images were observed during clear but non-photometric conditions.

Figure 1 shows $B$ and $J$ band images of the galaxy and an $R/B$
color map.  In the $B$ band image (Figure 1a) the classic figure 8 
shape of the R1-type ring can be seen.  The ring has some spiral
structure which causes the ring to be brighter on its north-east
and south-west sides than on its north-west and south-east sides.  
This means that the ring does not have perfect mirror symmetry about its
major axis.  The pinches near the bar ends can be more clearly
seen in the brighter north-east and south-west sides of the ring.

\section{THE FORM OF THE GRAVITATIONAL POTENTIAL }

To integrate orbits in the plane of the galaxy we require an estimate of
the gravitational potential.
We assume that the potential in the plane
$\Phi(r,\theta)$ is a sum of two components, an axisymmetric one, $\Phi_0(r)$, and
a component $\Phi_2(r)$ proportional to $\cos{2 \theta}$, 
\begin {equation}
\Phi(r,\theta) = \Phi_0(r) + \Phi_2(r) \cos(2 \theta)
\end{equation}
where $\theta$ is the azimuthal angle in the plane of the galaxy and $r$
is the radius.  
We take the bar to be aligned along the axis with $\theta=0$.
%Higher order Fourier components are neglected.

\subsection{The Axisymmetric Component}

The axisymmetric component
of the potential should be consistent with the rotation curve of the
galaxy at the location of the outer ring.
At the location of the outer ring in NGC~6782, dark matter is expected
to contribute to the rotation curve.  We therefore could not use
a potential derived solely from the light distribution as did
\cite{qui94} \& \cite{qui97} which were dynamical studies 
of bars in the central few kpc of galaxies.  

%Shape of rotation curves in ringed galaxies.
Unfortunately, there is no published rotation curve for NGC~6782. 
Few ringed galaxies have measured rotation curves at large radii.  However 
those few that have been observed, have rotation curves that are nearly flat.
For example, the rotation curve  
of ESO-509-98 is very flat (\cite{but96b}),
and the HI velocity field shows the rotation curve of NGC~3351 (of similar
morphological type) to be 
nearly flat (A. Bosma private communication).
Simulations of outer rings produce more realistic shaped rings
in model galaxies with flat rotation curves (\cite{byr94}).
In our orbit integrations, 
for the axisymmetric component of the potential, $\Phi_0$, 
we therefore assume a logarithmic form consistent with a flat rotation curve.
$\Phi_0$ is determined by one parameter, the circular velocity,
which we estimate from the Tully-Fisher relation (see below).  We
include scaling factors in all values which depend upon this velocity
so that when the actual circular velocity is known, these values 
can be corrected.

Near infrared images are superior to
visible images for dynamical studies because of their reduced
sensitivity to extinction from dust and because they are dominated by
light from an older cooler stellar population that is more
evenly distributed dynamically and a better
tracer of the stellar mass in the galaxy than the bluer,
hotter stars (e.g. \cite{fro88}; \cite{fro96}).
We use the $J$ band image to determine the gravitational potential
due to the luminous stellar component of the galaxy.
The $J$ band image was used because the sky is flatter outside the bar 
than it is in the $H$ band image and because it has higher signal
to noise than the $K$ band image.
The color $J-K = 1.0 \pm 0.05$ is constant across the bar, though 
the bulge of the galaxy ($r<7''$) is redder with $J-K = 1.06 \pm 0.05$.
The height of the rotation curve from the luminous stellar 
matter (traced in $J$ band) 
allows us to define a maximal disk, and show
that at the ring a significant dark component is needed to have a 
realistic flat rotation curve.

\subsection{Galaxy Inclination}
%inclination
Before the gravitational potential due to luminous stellar matter 
can be generated from the infrared image, 
we must correct for the inclination of the galaxy.  
A statistical study of R1$'$ ringed galaxies found
that these rings have observed axis ratios of 
$q_0 = 0.74 \pm 0.08$ and position angles on the sky with respect to the bar
of $\theta_0 = 90^\circ \pm 9^\circ$ (\cite{but95}).  \cite{but95}
found that R1$'$ rings are very nearly perpendicular
to the bar and are elongated.
Gas simulations of these rings support 
\cite{but95}'s finding for the ring alignment (\cite{byr94}).
We therefore assume that the ring is perpendicular to the bar.  This 
assumption reduces the degrees of freedom so that 
the major axis position angle is fixed by a choice for the inclination 
of the galaxy.

Another constraint on the galaxy inclination is obtained from the outermost
detected isophotes.  We constructed a 
sum of the $B,V$ and $R$ images weighted inversely by the noise in 
each band so as to maximize signal to noise in
the outer regions of the galaxy.  An outer isophote  is displayed
in Figure 1d.
This isophote has major axis oriented at 
a position angle of $\sim -45^\circ$ and has an axis ratio of $\sim 0.9$.
This suggests that the galaxy is not highly inclined. 
Since early-type galaxies are less often warped
than late-type galaxies (\cite{bos91}), it is unlikely that the galaxy
is warped at large radii. 
We therefore
corrected for the inclination, $i$ (where a face-on galaxy
has $i=0^\circ$), of the galaxy using various inclinations
and their accompanying position angles (see Table 1).  
Deprojected images of the galaxies  at these inclinations are
shown in Figure 2.

The inclination $i=41^\circ$ causes the ring in the plane of the galaxy
to be rounder or to have a larger axis ratio than for $i=35^\circ$.
Inclinations higher than $45^\circ$ cause the outer isophotes
of the galaxy (see Figure 1d)
to be very elliptical, or have an axis ratio smaller than $0.8$  
(see Table 1 for axis ratios).  These outer isophotes  
are not aligned with any feature in the galaxy so they should be close to 
circular.
Inclinations higher than $45^\circ$ also cause the ring
to be either aligned parallel to the bar or to be almost round 
(for $i=49^\circ$, the ring axis ratio is $\sim 1.0$).
This would be inconsistent with the statistics of 
of R1$'$ rings compiled by \cite{but95}.
For inclinations lower than $35^\circ$, the bar and the ring cannot 
be perpendicular. 
It is therefore unlikely that the inclination of the galaxy is
outside the range $35^\circ < i < 45^\circ$.

After correcting for inclination, the
gravitational potential in the plane of the galaxy  
traced by the luminous stellar matter 
was determined by convolving the $J$ image of NGC~6782 with a function
that depends on the vertical structure of the disk (\cite{qui94}).
Before convolution stars were removed from the $J$ image.  
The disk is assumed to have density
$\propto {\rm sech}(z/h)$ (following \cite{vdk88}) 
where $z$ is the height above the plane of the galaxy and
$h$ is the vertical scale height.
The resulting potential
is insensitive to the choice of vertical function
for functions such as sech, sech$^2$ and exponential with equivalent
$\langle z^2 \rangle$ (\cite{qui96}).
Since the galaxy is distant, a small vertical scale height was used 
$h=0.\hskip-2pt''5$.
\cite{qui96} found that doubling the vertical scale height results
in raising the $\Phi_2$ component of the gravitational potential by $\sim 10\%$. 
We have deliberately made $h$ small because the seeing in the images
causes artificial smoothing equivalent to increasing the size of $h$.

\subsection{What Do We Mean by a Maximal Disk?}

%Rotation curves for the axisymmetric component of the light
Figure 3 shows the rotation curve derived from the axisymmetric
component of the potential generated from the $J$ image for the different
galaxy inclinations.  The horizontal
line shown in Figure 3 is the circular rotational velocity computed using the 
Tully-Fisher Relation.  For an $H$ band total magnitude of $8.87$ measured from
our $H$ band image, we compute a circular velocity of $320$ km s$^{-1}$ using 
the relation given in \cite{pie92}.
All subsequent values given in this paper will be in units with respect
to this circular velocity.  
This circular velocity is also what we used for the flat rotation curve
axisymmetric component of the potential in our orbit integrations 
at the location of the ring.
The rotation curves shown in Figure 3 assume 
a distance of 50 Mpc ($H_0 = 75$ km s$^{-1}$ Mpc$^{-1}$) 
to the galaxy and a mass-to-light ratio of 
$M/L_J=1.23 \left({v_c \over 320 {\rm km~s}^{-1}}\right)^{2}
\left({50~{\rm Mpc}\over D}\right)$ 
or using the color of the bar
$M/L_K=0.69 \left({v_c \over 320 {\rm km~s}^{-1}}\right)^{2} 
\left({50~{\rm Mpc}\over D}\right)$ 
in solar units (see \cite{wor94}) 
where $L_J$ and $L_K$ are the luminosities in the $J$,
and $K$ bands and 
$v_c$ is the true (not yet measured) circular velocity of the galaxy.
% gunit *distance so that if the galaxy is really farther away
% then I need a lower m/l to get the right rot velocity
% if the galaxy is really nearer  then m/l must be higher
% if distance is nearer than rot velocity v_c is lower

For a ``maximal disk'' the rotation curve is attributed as
much as possible to be from the visible components so that
the halo could have a hollow core.
(Some authors call a ``maximal disk solution''
one with a smooth halo that extends into the nucleus of the galaxy).
The mass-to-light ratio listed above 
for the disk is what we take to give a ``maximal disk''.
This mass-to-light ratio was chosen so that the rotation curve
generated from the $J$ band light reaches above the circular velocity
predicted from the Tully-Fisher relation.
Once the true 
circular velocity for the galaxy has been observed, the mass-to-light ratio
for the maximal disk can be rescaled.    
Because the three dimensional nature of the bulge was not properly
taken into account in estimating the potential, the rotation
curve is higher (by $10-20\%$) than it should be in the central $0-20''$.
This is why we have chosen the mass-to-light ratio such that the rotation
curve is somewhat higher than the circular velocity near the galaxy nucleus
(see Figure 3).

We note that the rotation curve generated from the light
drops with increasing radius.  At the radius of the ring ($\sim 60-70''$
or $15-17$kpc),
a significant fraction of the mass must be from dark matter.
Matter outside of our image which we do not detect 
exerts a radial force outwards, so that the rotation curve predicted
from starlight should be even lower at large radii
than we show in Figure 3.

\subsection{The Non-Axisymmetric Component of the Gravitational Potential}

The non-axisymmetric component of the potential should be due solely to
the bar of the galaxy.  Since the bar is in the disk of the galaxy,
our inaccurate treatment of the bulge of the galaxy does not
affect our measurement for $\Phi_2(r)$ (defined in equation 1).
If luminous matter outside the image is axisymmetric
then once again, our estimate for $\Phi_2(r)$ is not affected by neglecting
this matter.  This means that our orbit integrations which use only
the $\Phi_2$ component derived from the luminous matter are not affected
by our inaccurate treatment of the bulge and outer disk.
Higher order Fourier components of the potential are neglected since
at the ring they are negligible.

The magnitude of the 
$\Phi_2$ component measured from the potential due to luminous matter is 
shown in Figure 4 for the various inclinations
assumed and for the maximal disk mass-to-light ratio
discussed above.  The $\Phi_2$ component drops off quickly with radius 
as expected for a quadrupolar potential term.
Also shown in Figure 4, an exponential function  
\begin {equation}
\Phi_2(r)= A \exp{(-r/a)},
\end{equation}
was fit to these $\Phi_2$ components.  
The numerical values for these fits are listed in Table 1.
These numbers show the strength of the non-axisymmetric component 
of the potential from the bar for the maximal disk, (corresponding to 
the rotation curves shown in Figure 3).
For the higher inclinations $\Phi_2$ component is substantially stronger 
because the bar
becomes longer once deprojected.  The exponential scale length 
of $\Phi_2$, $a$,  is also larger for the higher inclination case (see Table 1).
In the next section we discuss the effect of 
changing the disk mass-to-light ratio (and so the bar strength) 
on the morphology of the R1$'$ ring.

%$H_0 = 75$ km s$^{-1}$ Mpc$^{-1}$ 
%v = 3736 km/s -> 50Mpc pretty far away 
%1" = 250pc 

%As can be seen in from Byrd etal 1994, the slower the pattern speed, the larger
%the ring.  This pretty much just puts the ring at the OLR roughly.  So we 
%don't need to worry too much about what we choose as the pattern speed.
%Hints of dark matter:
%Buta 1987a tried a I band light predicted rotation curve, got a bump
%at the ring.  Either not much mass at the ring or dark matter dominated.
%In NGC 7351. Inner ring. so not so relevant.

\section{MODELING THE RING}

We integrate orbits in the plane of the galaxy for a gravitational
potential with axisymmetric component consistent with a flat rotation
curve and circular velocity determined from the Tully-Fisher relation.
The non-axisymmetric component of the potential is derived from exponential
fits to the $\Phi_2$ components generated from the $J$
band image for the various galaxy inclinations assumed.
In our integrations we vary the strength of the bar by adjusting
the mass-to-light ratio of the $J$ band image
and by keeping the circular velocity fixed.  
What we call the maximal disk corresponds to the mass-to-light
ratios for the rotation curves shown in Figure 3 and the
$\Phi_2$ components shown in Figure 4 with fitting parameters
listed in Table 1.  
Varying the mass-to-light ratio
of the disk corresponds to multiplying $\Phi_2$ by a constant
that is less than 1.
Since the maximal disk mass-to-light ratio is determined by our assumption
for the circular velocity (see discussion above), 
our results are not affected by
the fact that actual circular rotational velocity is not known.
Periodic orbits (or orbits that are closed in the frame in which
the bar is stationary) in the plane of NGC~6782 were found by numerical 
integration as in \cite{qui94}.  

%We first explore the morphology of these rings.

In Figure 5 we show periodic orbits near the OLR for the maximal disk
for a galaxy inclination of $41^\circ$.  
We see that the inner periodic orbits
are more pinched near the bar, and have a rounder appearance.
The outer orbits are more elongated and less pinched near the bar.
Points in Figure 5 
(and subsequent figures) are shown at equal timesteps along the orbit
so that the gas density in the orbit should be high in the pinches
near the bar.  Correspondingly the speed of the gas decreases in the pinches.
Measurement of the velocity field in the ring should constrain the degree
of cuspiness of the orbits.
As pointed out by \cite{kal91} the gas in the ring cannot be in an orbit
that intersects itself or that has loops.  We therefore only consider 
orbits that are are not self-intersecting.
%In his simulations \cite{sch81} found that the R1 rings correspond
%to periodic orbits with major axes that just touch the radius of the OLR.
%If we assume that this is true, the major axis of the ring
%sets the bar pattern speed. 
%That this happens is good evidence that the rotation curve of
%the galaxy is indeed nearly flat.
For the orbits shown in Figure 5a, the radius of corotation
is $37.7''$ and the bar angular rotation rate or pattern speed is 
$35.0$ Gyr$^{-1} \times 
\left({v_c \over 320 {\rm km~s}^{-1}}\right)
\left({50~{\rm Mpc}\over D}\right)$. 
% 41'' with wrong pixel scale
For our rotation curve, 
this pattern speed places the radius of corotation just past the end of the bar
as predicted theoretically and inferred from observations of bars.
Figure 5 shows that the maximal disk provides a good fit to
the morphology of the ring for a galaxy inclination of $i=41^\circ$.
In the following sections we explore the sensitivity of the ring
morphology to the bar strength, the bar pattern speed and 
the galaxy inclination.

\subsection{Varying the Strength and Pattern Speed of the Bar}

Figure 6 shows comparisons between 
periodic orbits with the same apogee for different bar strengths
and pattern speeds at a galaxy inclination of $i=41^\circ$.  
All figures compare an orbit shown in Figure 5a that has a maximal disk 
mass-to-light ratio with a similar orbit for either lower bar mass-to-light
(Figure 6a), faster bar (Figure 6b) or both (Figure 6c).
Figure 6a shows the comparison for two bars with the same pattern speed. 
We note that the weaker bars have rounder less pinched orbits of
the same apogee.
Figure 6b a comparison between a slow and a faster bar with the same
bar strengths.   We can see that the faster bar has a more elongated 
and strongly pinched orbit of the same apogee.  
In Figure 6c we can see that by decreasing the bar strength
and increasing the bar pattern speed, differences 
in the shapes of orbits with the same apogee
can be minimized.

%xxxxx
Weakening the bar potential at the location 
of the ring (equivalent to decreasing the disk mass-to-light ratio) 
to less than $70\%$ of the maximal disk value causes the 
pinched orbits to be rounder and less pinched than the observed ring.
We find that it is not possible to consistently match the morphology
of the ring by raising the bar pattern speed in a weaker bar, 
since this decreases the radius of corotation to within
the bar ends.  For the orbits
shown in Figure 5 (and the solid line in Figure 6abc) the corotation
radius lies just outside the end of the bar at a radius
of $37.7''$.  The faster bar (with
corresponding orbit shown as open points in Figure 6c) 
which would fit the ring morphology
places the radius of corotation at $33.1''$ which
lies within the end of the bar, a situation which is not thought
to be theoretically possible (\cite{con89b}).
As a result we find that for an assumed inclination of $i=41^\circ$
the bar mass-to-light ratio must be greater than 70\% of the maximal disk value.

\subsection{Changing the Galaxy Inclination}

For a lower galaxy inclination of $i=35^\circ$ the strength 
of the $\Phi_2$ component is about half
as large of that with $i=41^\circ$
at the location of the ring (see Figure 4).
As expected from the previous section it is not possible to match
the ring morphology with the periodic orbits
without increasing the mass-to-light ratio
past what we have defined as the maximal disk value or
decreasing the radius of corotation to a radius smaller than the bar length. 
Figure 7 shows orbits integrated for a maximal disk mass-to-light
ratio with a bar corotation radius of $32.2''$.    
Orbits with larger apogees
than shown in Figure 7 become self-intersecting with small loops at their
minor axes and are probably unstable.
The non-self-intersecting periodic orbits are too round to match 
the observed ring morphology.
It is not possible to resolve the problem by raising the bar angular
rotation rate since this would place the radius of corotation within
the end of the bar.
%41.8''=30.6Gyr with wrong pixel scale

It would be possible to have a stronger bar or a larger
mass-to-light ratio with a more carefully estimated maximal disk value
at $i=35^\circ$.
For example  if the rotation curve of the galaxy decreases with radius
%or if the vertical scale height of the disk is lower 
near the ring then the maximal disk
value for mass-to-light ratio could be higher.
The ring enveloping the bar which is quite blue (see Figure 1c)
may contain a large gas mass.  Including this gas mass increase the strength
of the non-axisymmetric component ($\Phi_2$) of the potential and so
cause the periodic orbits near the OLR to be more elongated.
%In short we find that a mass-to-light ratio larger than the one
%assumed here is required to match the morphology of the ring
%at a galaxy inclination of $35^\circ$.
In short we find that for an assumed inclination of $i=35^\circ$
the bar mass-to-light ratio must be greater than the maximal disk value
assumed here.

%Here we discuss the possibility that the galaxy inclination is
%higher than $41^\circ$.  
For higher galaxy inclinations, the outer ring becomes rounder and the
bar lengthens  (see Figure 2). 
A mass-to-light ratio lower than the maximal disk value is required 
to match the observed morphology of the ring.
In Figure 8 we show periodic orbits for a galaxy
inclination of $45^\circ$ that resemble
the morphology of the ring for a mass-to-light
ratio that is $75\%$ of the maximal disk value.
Since the average radius of the ring is larger for
this galaxy inclination than for lower inclinations,
the bar angular rotation rate must be lower.
%The bar angular rotation
%rate for the orbits shown in Figure 8a is 
%$29.0$ Gyr$^{-1}  \times \left({v_c \over 320 {\rm km~s}^{-1}}\right)$
%and the radius of corotation is $44.2''$, again outside
%the end of the bar as expected.
Mass-to-light ratios lower than $60\%$ 
again cause the ring to be too round to match the morphology of the ring.
Higher mass-to-light ratios than $90\%$ eliminated the 
R1 periodic orbits (the resonance was very strong)
at the radius of the ring for pattern speeds that 
kept the corotation radius outside the bar ends.
As a result we find that for an assumed inclination of $i=45^\circ$
the bar mass-to-light ratio must be within $75 \pm 15 \%$ 
of the maximal disk value.

It is extremely unlikely that the galaxy inclination is much higher
than $45^\circ$ since at $50^\circ$ the ring is almost round
and the outer isophotes of the galaxy are even more elongated
(see Figure 2).

%\subsection{Other effects}

%Including gas mass in the rotation curve should cause stronger
%limits on the mass-to-light ratios.

%The effect of the scale length.  Lowering the vertical scale length causes
%the bar to be stronger.  Since we have chosen a pretty small
%value already so our results should be unaffected.  Choosing large
%scale height lowers $\Phi_2$ and so causes higher mass-to-light
%ratios to be required. 

% Phi_4 drops off way too fast

\section {SUMMARY AND DISCUSSION}

\subsection {Underlying Assumptions}
We have assumed the following in modeling the R1$'$ outer ring in 
NGC~6782:

\noindent
1) The ring morphology consists of gas in periodic non-self intersecting orbits
near the Outer Lindblad Resonance, and 
spiral structure in the ring does not cause the morphology
to deviate significantly from these periodic orbits.

\noindent
2) The bar is perpendicular to the ring. This allowed us to determine
the position angle for a given galaxy inclination.

\noindent
3) The rotation curve is flat.

\noindent
4) The $\Phi_2$ non-axisymmetric component 
of the potential is only due to the bar as seen in the $J$ band image
(gas is neglected) and does not twist (equation 1).

\noindent
5) The maximal disk mass-to-light ratio is well estimated from
the axisymmetric component of the $J$ band generated potential.

Many of these assumptions can be constrained with a velocity
field that can determine the inclination and measure the rotation
curve of the galaxy.  However the first assumption listed above is
of particular concern.  Strong spiral structure in the ring will 
cause the gas to deviate from the periodic orbit families explored
here.  While it is not unreasonable to expect that the gas
is close to the periodic orbits (in the same way gas in the Milky
way is primarily undergoing circular motion despite its spiral
structure), future work should both study R1 ring galaxies with
minimal spiral structure and investigate the role of the spiral structure
in these rings.

\subsection {Summary and Discussion}
In this paper we have 
explored the shape of the periodic orbits near the Outer Lindblad
resonance in a ring galaxy
using a non-axisymmetric gravitational potential 
based upon a near-infrared image of the bar.
We find that the shape of the non-self-intersecting
periodic orbits at the OLR is affected by the strength of the non-axisymmetric
component $\Phi_2$ of the gravitational potential. 
%but is insensitive to its exponential scale length.  
A stronger bar (corresponding
to a larger $\Phi_2$) or a faster bar result
in more elongated orbits at the radius of the ring.    

Using the above assumptions, and  
comparing outer ring morphology of NGC~6782 with the 
integrated non-self-intersecting periodic orbits
we find that the bar mass-to-light ratio can be constrained
given an assumed galaxy inclination.  
For a galaxy inclination
of $41^\circ$ we find that a bar mass-to-light ratio greater than $70\%$ of 
the maximal disk value is needed to match the ring morphology.
It is not possible to match the morphology of the ring
with a weaker and faster bar since this places the bar corotation radius
within the end of the bar.
For $i=45^\circ$ we find a mass-to-light ratio of $75\pm 15\%$
of the maximal disk value matches the morphology of the ring.
For $i=35^\circ$ a value larger than the maximal disk value assumed here 
is required.
A larger mass-to-light ratio could be allowed if a large gas mass is found 
in the ring enveloping the bar (increasing the $\Phi_2$ component of the
potential), or if the rotation curve decreases near the ring.

Larger galaxy inclinations are unlikely for the following
reasons:  1) The ring becomes round or aligned with the bar which 
is not supported by statistics of R1-type rings
(\cite{but95}). However, NGC~6782 could be a special case.  
2) The outer isophotes of the galaxy become significantly
elongated.
Deeper images showing the shapes of the isophotes past the ring
may help to constrain the inclination angle of the galaxy.

We note that here that the method considered here places a constraint on 
the the strength of the non-axisymmetric ($m=2$) component
of the potential from the bar.  Since the bar is necessarily 
a disk component this
leads directly to a constraint on the disk mass-to-light ratio.
Once the rotation curve is observed the morphology of 
the ring gives a constraint on the disk mass-to-light
ratio which is independent of any assumptions about the halo
or dark matter distribution.
As a result a measured rotation curve would allow us to measure
the core radius of the dark matter halo using our values for
the mass-to-light ratio.
A measured rotation curve will also enable
us to place a value on the maximal disk mass-to-light ratio 
assumed here and check whether
our assumption of a flat rotation curve is a good one.

Modeling of an observed velocity field in the ring will make it
possible to measure the mass-to-light ratio of the bar with
more precision than with a purely morphological comparison
as done here.  Velocities observed along the major axis of the ring should
constrain the inclination of the galaxy.
Highly pinched orbits have slow speeds in their pinches, as inferred
from small spacing in the equal time step points shown in Figures 5-8.
The velocity field should also therefore limit 
which particular orbits are represented in the ring.
The asymmetry of the velocity field 
will also constrain the degree of deviation 
from the periodic orbits caused by spiral structure in the ring.

We also plan to observe  
other ring galaxies to find if high mass-to-light ratios 
are required generally.
Modeling of galaxies with different orientations should
help resolve the uncertainties caused by projection.

Gas simulations of rings could be studied to discover how well
closed orbits match the gas morphology in these systems, which
particular orbits collect gas, and how close outer rings are to being
perpendicular to their bars.  The effect of spiral structure on the
morphology of the ring should also be studied.

If a near maximal disk value for the bar mass-to-light ratio is 
indeed required (as suggested here)
then either the inner parts of galaxies have little dark matter,
or the dark matter contained in the disk of the galaxy
is non-axisymmetric and rotates with the bar.
The second possibility implies that 
the ``conspiracy of shapes'' suggested by \cite{sac94} extends
into the bar, and would lend support to the idea that 
dark matter halos are flattened (\cite{sac94}; \cite{oll96}), since if
the dark matter rotates, it should be flattened.
%In other words, if there is a significant percentage
%of dark matter present in the inner disk of the galaxy, then it must
%be rotating with the bar.  

%Some previous work has found that dark matter halos are not spherical.
%For example \cite{sac94} found studying a polar ring galaxy E6,E7
%and \cite{oll96} with careful modeling of HI velocity profiles in edge on
%galaxies.  However previous studies of the flaring of the HI disk
%in the milky way suggest that our galaxy has a spherical halo?
%(reference)
%For studies of our galaxy, Chris Flynn finds that the local milky
%way is 75 percent not dark? But is this still way below what is needed
%for the rotation curve?

\acknowledgments

We acknowledge helpful discussions and correspondence 
with R. Buta, A. Gould, D. Weinberg, K. Griest, P. Sackett, G. Rieke 
\& A. Bosma.
We thank the referee for helpful comments which significantly improved
this paper.
The OSU galaxy survey is being supported in part by NSF grant AST 92-17716.
%OSIRIS was built with substantial aid from NSF grants AST 90-16112 and AST 92-18449.
A.C.Q. acknowledges the support of a Columbus fellowship and a
grant for visitors from L'Observatoire de Marseille.
J.A.F. thanks
Roger Davies for his hospitality at Durham University and PPARC for
partial support via a Visiting Senior Research Fellowship.

\clearpage

\begin{figure*}
\caption[junk]{
a) Grayscale $B$ band image of NGC~6782 overlayed with 
contours of the bar 0.5 mag apart.  The image
is uncalibrated so we do not know the absolute scale.
b) $J$ band contours of the bar. 
The brightest contour is at $15.5$ mag/arcsec$^2$ and the
difference between contours is $0.5$ mag/arcsec$^2$.
c) B/R color map similar to that shown by Buta 1995.
d) Low surface brightness contour in an image that is a noise weighted
sum of the $B,V$ and $R$ band images.   Note the change in angular scale between
this figure and the other figures.  We note that this outer isophote
is almost round suggesting that the galaxy is not highly inclined.
\label{fig:fig1} }
\end{figure*}

\begin{figure*}
\caption[junk]{
Deprojected $B$ band images for the inclinations and position angles
listed in Table 1.  Note that the higher the galaxy inclination
the rounder the ring and the longer the bar.  The position angles
have been chosen so that the bar is approximately perpendicular to the ring.
These images have been rotated so that the bar has a major
axis at $PA \approx 90^\circ$ which is the same as that observed in the
original image.
a) $i=35^\circ$.
b) Same as a) but for $i=41^\circ$.
c) Same as a) but for $i=45^\circ$.
d) Same as a) but for $i=49^\circ$.
\label{fig:fig2} }
\end{figure*}

\begin{figure*}
\caption[junk]{
The rotation curves from the axisymmetric component of
the $J$ band generated gravitational potential 
using the mass-to-light ratio described in the text as `maximal disk'.
Rotation curves for galaxy inclinations of $i=35^\circ$
and $45^\circ$ have been plotted as solid
lines.  There is little difference between them.
The horizontal dotted line represents a flat rotation curve with a
circular velocity of $320$ km/s predicted using the Tully-Fisher relation.
Once the true circular velocity, $v_c$,
of the galaxy is observed, the mass-to-light
ratio for the maximal disk should be rescaled.
For a non-maximal disk the rotation curve resulting
from the luminous stellar matter would be lower than that shown here.
\label{fig:fig3} }
\end{figure*}

\begin{figure*}
\caption[junk]{
The amplitude of the non-axisymmetric component $\Phi_2$ from the bar 
of the $J$ band generated gravitational potential for the maximal disk.
The solid lines are for galaxy inclinations of 
$i=45^\circ$, $41^\circ$ and $35^\circ$ in order of 
decreasing height.
Once the true circular velocity, $v_c$,
of the galaxy is observed, the mass-to-light
ratio for the maximal disk should be rescaled.
The dotted lines are exponential fits to these curves with strengths
and scale lengths listed in Table 1.  
$\Phi_2$ is substantially weaker for the lower galaxy inclinations.
\label{fig:fig4} }
\end{figure*}

\begin{figure*}
\caption[junk]{
Comparison of periodic orbits with the morphology of the ring at a
galaxy inclination of $i=41^\circ$.
a) Grayscale of the deprojected galaxy.
b) Periodic orbits near the Outer Lindblad Resonance.
The mass-to-light ratio of the bar in units of percent
of the maximal disk value (MD) and the corotation radius ($r_{cr}$)
are printed in the right hand corner of the plot.   
Points are plotted at equal timesteps in the rotating 
frame in which the bar is still.
Note that speeds are slower in the pinches near the bar ends.
The dotted circle shows the location of the Outer Lindblad Resonance.
The maximal disk provides a good representation for the morphology
of the ring.
\label{fig:fig5} }
\end{figure*}

\begin{figure*}
\caption[junk]{
a)  The effect of varying the bar strength on the shape of the orbits
with the same apogee for a galaxy inclination $i=41^\circ$.
A periodic orbit for the maximal disk (solid line) 
compared to a similar orbit for weaker bars with
mass-to-light ratios that are $70\%$ and $50\%$ 
as large as the maximal disk value
(open points).  These orbits are derived from bars with the
same corotation radii.  
The dotted circle shows the location of the Outer Lindblad Resonance.
Bar strengths, in units of percent of the maximal disk value (MD), and
the corotation radii ($r_{cr})$ are printed in the upper right hand corners of
the plots.
b)  The effect of varying the bar pattern speed on the shape of the orbits
with the same apogee.
A periodic orbit for the maximal disk (solid line) 
compared to a similar orbit with the same bar mass-to-light ratio but
a faster bar.  
c) By increasing the speed and decreasing the strength of the bar 
differences in the orbit shapes can be minimized.
A pinched non-self-intersecting orbit for the  
maximal disk (solid line) 
compared to a similar orbit (open points) 
for a faster and weaker bar.
\label{fig:fig6} }
\end{figure*}

\begin{figure*}
\caption[junk]{
Comparison of periodic orbits with the morphology of the ring at a
galaxy inclination of $i=35^\circ$.
a) Grayscale of the deprojected galaxy.
b) Periodic orbits near the Outer Lindblad Resonance.
The mass-to-light ratio of the bar in units of percent
of the maximal disk value (MD) and the corotation radius ($r_{cr}$)
are printed in the right hand corner of the plot.   
Points are plotted at equal timesteps in the rotating 
frame in which the bar is still.
The dotted circle shows the location of the Outer Lindblad Resonance.
%c) Same as b) but with a faster bar. 
The maximal disk value for the mass-to-light ratio 
produces non-self-intersecting orbits 
that are insufficiently elongated
to be consistent with the morphology of the ring.
\label{fig:fig7} }
\end{figure*}

\begin{figure*}
\caption[junk]{
Comparison of periodic orbits with the morphology of the ring at a
galaxy inclination of $i=45^\circ$.
a) Grayscale of the deprojected galaxy.
b) Periodic orbits near the Outer Lindblad Resonance.
The mass-to-light ratio of the bar in units of percent
of the maximal disk value (MD) and the corotation radius ($r_{cr}$)
are printed in the right hand corner of the plot.   
Points are plotted at equal timesteps in the rotating 
frame in which the bar is still.
The dotted circle shows the location of the Outer Lindblad Resonance.
The mass-to-light ratio of $75\%$ of the maximal disk value 
provides a good representation for the morphology of the ring.
c) Same as b) but with a weaker bar. 
\label{fig:fig8} }
\end{figure*}

\vfill

\clearpage

% table1.tex
\begin{deluxetable}{crrrrrrrr}
%\footnotesize
\scriptsize
\tablewidth{0pt}
\tablecaption{Galaxy Orientation and Exponential Fits to $\Phi_2$  \label{tbl-1}}
\tablehead{ 
\colhead{Inclination} & 
\colhead{PA$^i$} & 
\colhead{$A$(km s$^{-1}$)$^2$$^j$} & 
\colhead{$a$(arcsec)$^k$} & 
\colhead{Outer Axis Ratio$^l$}} 
\startdata
$35^\circ$ & $-58^\circ$ & $3.78 \times 10^4$  & $17.5$ & $1.0$ \cr
$41^\circ$ & $-45^\circ$ & $3.51 \times 10^4$  & $22.5$ & $0.83$ \cr
$45^\circ$ & $-32^\circ$ & $3.63 \times 10^4$  & $24.3$ & $0.76$ \cr
$49^\circ$ & $-25^\circ$ & $3.40 \times 10^4$  & $26.3$ & $0.70$ \cr
% wrong pixel scale below
%$35^\circ$ & $58^\circ$ & $3.78 \times 10^4$  & $19.0$ & $1.0$ \cr
%$41^\circ$ & $45^\circ$ & $3.51 \times 10^4$  & $24.4$ & $0.83$ \cr
%$45^\circ$ & $32^\circ$ & $3.63 \times 10^4$  & $26.4$ & $0.76$ \cr
%$49^\circ$ & $25^\circ$ & $3.40 \times 10^4$  & $28.6$ & $0.70$ \cr
\enddata
\tablenotetext{i}{Galaxy major axis Position Angle required for
the bar to be perpendicular to the ring at the given inclination.}
\tablenotetext{j}{For a maximal disk mass-to-light ratio (see
equation 2).  Once the true circular velocity, $v_c$,  of the galaxy is known
these numbers should be multiplied by 
$\left({v_c \over 320 {\rm km~s}^{-1}}\right)^2$.}
\tablenotetext{k}{Exponential scale length of $\Phi_2$ (see equation 2).}
\tablenotetext{l}{Axis ratio of the deprojected 
outer isophote shown in Figure 1d.}
\end{deluxetable}

\end{document}